\documentclass[preprint,5p,twocolumn]{elsarticle}
\usepackage{slashed} 

\def\eq#1\en{\begin{equation} #1 \end{equation}} 

\def\eqa#1\ena{\begin{eqnarray} #1 \end{eqnarray}}

\usepackage{graphicx,color,amsmath,amssymb,hyperref,epsfig,tensor,epstopdf}

\makeatletter
\newcommand{\fmslash}[2][0mu]{%
  \mathchoice
    {\fmsl@sh\displaystyle{#1}{#2}}%
    {\fmsl@sh\textstyle{#1}{#2}}%
    {\fmsl@sh\scriptstyle{#1}{#2}}%
    {\fmsl@sh\scriptscriptstyle{#1}{#2}}}
\newcommand{\fmsl@sh}[3]{%
  \m@th\ooalign{$\hfil#1\mkern#2/\hfil$\crcr$#1#3$}}

\journal{Physics Letters B}

\begin{document}

\begin{frontmatter}

\title{Spacetime Deformation Effect on the Early Universe \\and the PTOLEMY Experiment}

\author{Raul Horvat} \address{Rudjer Bo\v skovi\' c Institute, Experimental
Physics Division, Bijeni\v{c}ka 54 Zagreb, Croatia}
\author{Josip Trampetic}
\address{Rudjer Bo\v skovi\' c Institute, Experimental
Physics Division, Bijeni\v{c}ka 54 Zagreb, Croatia}
\address{Max-Planck-Institut f\"ur Physik,
(Werner-Heisenberg-Institut), F\"ohringer Ring 6, D-80805 M\"unchen,
Germany}
\author{Jiangyang You}
\address{Rudjer Bo\v skovi\' c Institute, Theoretical Physics Division, Bijeni\v{c}ka 54 Zagreb, Croatia}


\begin{abstract}
Using a fully-fledged formulation of gauge field theory
deformed by the spacetime noncommutativity, we study its impact on relic
neutrino direct detection, as proposed recently by the PTOLEMY experiment.
The noncommutative background tends to influence the propagating neutrinos
by providing them with a tree-level vector-like coupling to photons, enabling thus otherwise  right-handed (RH) neutrinos to be thermally produced in the early
universe.  Such a new component in the universe's background radiation has been switched today
to the almost fully active sea of non-relativistic neutrinos, exerting consequently some
impact on the capture on tritium at PTOLEMY.  The peculiarities of our nonperturbative
approach tend to reflect in the cosmology as well,
upon the appearances of the coupling temperature, above which
 RH neutrinos stay permanently decoupled from thermal environment. This entails the maximal scale of noncommutativity as well, being of order of $10^{-4} M_{Pl}$, above which there is no impact whatsoever on the capture rates at PTOLEMY.
The latter represents an exceptional upper bound on the scale of noncommutativity coming from
phenomenology.
\end{abstract}


\end{frontmatter}

Out of the three pillars of the standard Big Bang model, Big Bang
nucleosynthesis (BBN) \cite{Sarkar:1996} relates directly to neutrinos and
provides us with useful (but somewhat indirect) information about the
universe when it was just about 1 minute old. Another pillar of the Big Bang, the cosmic microwave background radiation (CMBR), the relic radiation left over from the moment the universe cooled off and became transparent, allows us to see directly into cosmos when it was 380000 years old. It was  measured recently so precisely that this has deepened our understanding of the early universe to a hitherto undreamed-of scale \cite{Ade:2015xua}.  A related prediction of the standard theory is the undisputed existence of a relic neutrino background, whose
direct detection would enable to see what was the universe doing when it was
only about one second old.

Given the fact that neutrinos interact only feebly with ordinary matter, the relic neutrino background turns out to be composed altogether of neutrinos which are nonrelativistic today, making them consequently very difficult to directly detect in the laboratory.  This also turns out to be the only primary source of nonrelativistic neutrinos in the universe at present.

A first promising proposal to detect such a cold sea of neutrinos at the
temperature of around 2 Kelvin, was to use the inverse beta decay of Tritium nucleus, $\nu_{e} \,+\, ^3$H$\:\rightarrow {^3}$He$ \;+\; e^{-}$ \cite{Weinberg:1962zza}.  The possibility of detecting such a background experimentally, using this
process, was investigated in \cite{Irvine:1983nr}.  Earlier attempts to detect relic neutrino sea were precisely compiled, but also strongly criticized in \cite{Langacker:1982ih}.  With the recently proposed PTOLEMY experiment, with an energy resolution $\Delta \sim 0.15$ eV and implementing a 100 gram sample of Tritium, the detection of relic neutrino background might soon become a dream come true \cite{Betts:2013uya}.

For a long time, BBN has proven as one of the most powerful available probes
of physics beyond the standard model (SM), giving many interesting
constraints on particle properties.  The BBN has played a central role in
constraining particle properties since the seminal paper of Steigman,
Schramm and Gunn \cite{Steigman:1977kc}, in which the observation-based
determination of the primordial abundance of $^{4}$He was used for the first
time to constrain the number of light neutrino species.  Later, with the
inclusion of other light element abundances (D,$^{3}$He and $^{7}$Li) and
their successful agreement with the theoretically predicted abundances, many
aspects of physics beyond SM have been further constrained
\cite{Malaney:1993ah}.  One usually parameterizes the energy density of new
relativistic particles in the early universe in terms of the effective
additional number of neutrino species, $\Delta N_{eff}$.  After decades in
which $\Delta N_{eff}$ remaind poorly constrained, a combination of $Planck$
observations ($Planck$ 2015 results \cite{Ade:2015xua}) with other
astrophysical data has recently strongly constrained the neutrino sector of
the theory, giving $(\Delta N_{eff})^{max} = 0.33$.  Since the data favour
$N_{eff} = 3.15 \pm 0.23$ \cite{Ade:2015xua}, one finds this consistent with
the standard model value $N_{eff} = 3.046$ itself.

Entertaining the possibility to thermally produce right-handed (RH) neutrinos $\nu_R$ in some extension of the standard model, we note that the energy density of 3 light RH neutrinos is equivalent to the effective number $\Delta N_{eff}$ of additional doublet neutrinos
\begin{equation}
\Delta N_{\nu} = 3 \left
(\frac{T_{\nu_{R}}}{T_{\nu_{L}}} \right )^4 ,
 \label{eq1}
 \end{equation}
where $T_{\nu_{L}}$ is the temperature of the SM neutrinos, being the same
as that of photons down to $T \sim 1$ MeV.  Hence we have
\begin{equation} 3
\left (\frac{T_{\nu_{R}}}{T_{\nu_{L}}} \right )^4 \lesssim {(\Delta N_{eff})^{max}}.  \label{eq2}
\end{equation}
In the following we take the latest Planck result, $(\Delta N_{eff})^{max} =0.33$.

How the temperature of $\nu_{R}$'s, which decoupled at $T_{dec}$, relates to
the temperature of still interacting $\nu_{L}$'s below $T_{dec}$, stems
easily from the fact that the entropy in the decoupled species and the
entropy in the still interacting ones are separately conserved.  The ratio
of the temperatures is a function of $T_{dec}$ and is given by
\cite{Srednicki:1988ce,Gondolo:1990dk}
\begin{equation}
\frac{T_{\nu_{R}}}{T_{\nu_{L}}} = \left [\frac{g_{*\nu_R }(T_{dec})}{g_{*\nu_R }(T_{\nu_L})} \frac{g_{*S}(T_{\nu_L})}{g_{*S}(T_{dec})} \right ]^{1/3} ,
\label{eq3}
\end{equation}
where $g_{*\nu_R }$ and $g_{*S}$ are the degrees of freedom specifying the entropy of the decoupled and of the interacting species, respectively \cite{Sarkar:1996,Kolb}.  Since in our case we ignore the possibility that the decoupled particles may subsequently annihilate into other non-interacting species, $g_{*\nu_R }$ stays constant after decoupling and therefore, for all practical purposes, the first ratio in (\ref{eq3}) equals unity.

Now, combining (\ref{eq2}) with (\ref{eq3}) and noting that at the time
of BBN $g_{*S}(T_{\nu_L}\sim MeV) = 10.56$~\cite{Husdal:2016haj}, one arrives at
\begin{equation}
g_{*S}(T_{dec}) \gtrsim \; \frac{24.1}{(\Delta
N^{max}_{\nu})^{3/4}}.
 \label{eq4}
 \end{equation}
With the latest bound $\Delta N^{max}_{\nu} =0.33$, (\ref{eq4}) implies $g_{*S}(T_{dec}) > 55.3$ which, given the temperature dependence of $g_{*S}$~\cite{Husdal:2016haj}, can be seen to enforce $T_{dec} \gtrsim T_C $, where $T_C$ is the critical temperature for the
deconfinement restoration phase transition, $T_C \sim 200$ MeV.

Since background neutrinos are ultra-relativistic at freeze out, the
left-handed neutrinos $\nu_L$ almost exactly coincide with the
left-helical neutrinos $\nu_l$ (similarly for anti-neutrinos), which means
that in the standard theory the right-helical neutrinos $\nu_r$ are
(practically) not populated at all.  If, by some mechanism, the right-handed
neutrinos $\nu_R$ were thermally produced in the early universe, they again
almost exactly coincide with the right-helical neutrinos $\nu_r$ (and
similarly for anti-neutrinos).  Since for free-streaming neutrinos it is
their helicity that is conserved \cite{Duda:2001hd}, and the relic neutrino
background is non-relativistic today (for neutrino masses $m_{\nu} \gtrsim
10^{-3}$ eV), one finds that non-relativistic right-helical neutrinos
$\nu_r$ are no longer inert, in fact, they can (almost) equally be captured
in the $\nu_{e} \,+\, ^3$H$\:\rightarrow {^3}$He$ \;+\; e^{-}$ process as
their left-helical partners $\nu_l$'s do.

As calculated in detail in \cite{Long:2014zva} (for earlier calculations see
also \cite{Cocco:2007za}) the total capture rate boils down to a simple
expression
\begin{equation}
\Gamma = \bar{\sigma} [n(\nu_l) + n(\nu_r)] N_{trit} ,
\label{eq5}
\end{equation}
where $\bar{\sigma} \approx 4 \times 10^{-45}$ cm$^2$, $N_{trit}$ is the number of tritium nuclei and $n(\nu_l)$ and $n(\nu_r)$ are the number densities of left- and right-helical neutrinos per degree of freedom.  In the standard theory, both active degrees of freedom for the massive Majorana case equally contribute to the process, while in the Dirac case only one active (out of four) degrees of freedom
does so.  Hence, the capture rate in the Majorana case is twice that in the
Dirac case \cite{Long:2014zva}.

Note that the thermal  production of right-handed Dirac neutrinos in the
early Universe has been discussed before in the literature and the
cosmological bound on the extra effective number of neutrino species can be
satisfied \cite{Zhang:2015wua}.  A possible way to discriminate between
thermal and non thermal cosmic relic neutrinos was proposed in
\cite{Huang:2016qmh}.

\begin{figure}[t]
\begin{center}
\epsfig{file=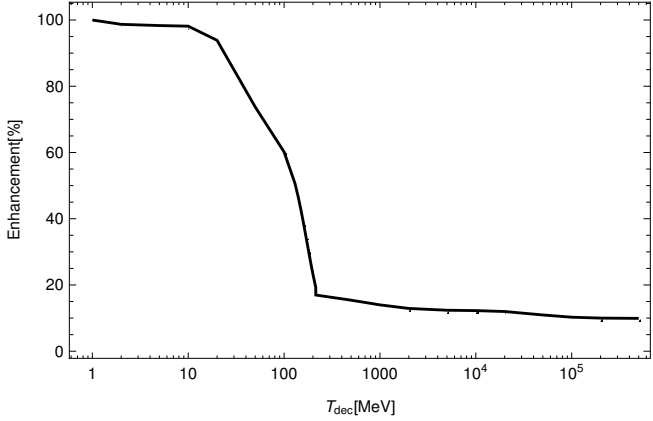,width=8.5cm}
\end{center}
\caption{PTOLEMY capture rate enhancement (\%) versus $T_{\rm dec}$, based on the temperature dependence of $g_{*S}$ given in~\cite{Husdal:2016haj}.}
\label{fig:Enhancement}
\end{figure}

When the right-handed neutrinos are produced by some non-standard mechanism
in the early universe, their relative contribution in (\ref{eq5}) is given
by the ratio of the temperatures cubed ($n_{\nu} \sim T_{\nu}^3$), as given
by (\ref{eq3}).  This is because the ratio (\ref{eq3}) remains constant
below $T \sim$ MeV, as both $\nu_l$ and $\nu_r$ are then decoupled.  This
implies around 20\% magnification of the capture rate at PTOLEMY if $T_{dec}
\sim T_C$, and around 10\% magnification if $T_{dec} \sim T_{EW}$, where
$T_{EW}$ is the critical temperature for electroweak phase transition,
$T_{EW} \approx 200$ GeV. We plot the capture rate enhancement (percentage) versus decoupling temperature in Fig. \ref{fig:Enhancement}.

As a working example to realize a thermal production of right-handed neutrinos
$\nu_R$ in the early universe, via plasmon decay into neutrino pairs \cite{Adams:1963zzb,Bandyopadhyay:1969ck,Grasso:1993bp}, we propose a fully-fledged Seiberg-Witten (SW) map based  \cite{Seiberg:1999vs,Schupp:2008fs} $\theta$-exact formulation of noncommutative (NC) gauge field theory. This model further preserves unitarity \cite{Aharony:2000gz}, has a correct commutative limit \cite{Horvat:2011iv,Horvat:2011qn,Horvat:2009cm,Horvat:2011wh,Horvat:2010sr,Horvat:2012vn}, and for which it has been shown that a nice UV/IR behavior at the quantum level can in fact  be achieved, especially when supersymmetry is included \cite{Horvat:2011bs,Horvat:2011qg,Horvat:2013rga,Trampetic:2015zma,Martin:2016zon,Martin:2016hji,Martin:2016saw}.

Alluding to the above model, we now introduce an effective coupling
involving neutrinos and photons on NC spaces which can result in thermal
production of right-handed neutrinos in the early universe, giving
consequently a nonzero right-helicity component in the cosmic neutrino background.  Such an additional component would result in an enhancement to the Tritium capture rate in the PTOLEMY experiment, which, if observed and assuming to be due to the space-time noncommutativity \cite{Calmet:2001na,Chaichian:2001py,Schupp:2002up,Minkowski:2003jg}, could potentially probe its associated scale.

In the presence of space-time noncommutativity it is possible to directly couple neutrinos  to Abelian gauge bosons (photons) via a star($\star$)-commutator in the NC covariant derivative of the action \cite{Schupp:2002up}:
\begin{eqnarray}
S&=&\int -\frac{1}{4}\widehat F^{\mu\nu}\widehat
F_{\mu\nu}+i\bar{\widehat\Psi} \big(\slashed{\widehat D}-m\big)\widehat\Psi,
\label{action}\\
\widehat D^\mu&=&\partial^\mu\phantom{..}-ie\kappa[\widehat
A^\mu\stackrel{\star}{,}\phantom{...}],
\label{hatD}
 \end{eqnarray}
with $\widehat A^\mu, \widehat\Psi$ being noncommutative fields on the Moyal space and a coupling constant $e\kappa$ corresponds to a multiple (or fraction) $\kappa$ of the positron charge $e$. The $\star$-product above is associative but, in general, not commutative - otherwise the proposed coupling to the noncommutative photon field $\widehat A_\mu$ would of course be zero.

In view of the NC covariant derivative $\widehat D^\mu$ (\ref{hatD})
one may think of the noncommutative neutrino field $\widehat\Psi$ as having left charge $+e\kappa$, right charge $-e\kappa$ and total charge zero.  From the perspective of non-Abelian gauge theory, one could also say that the neutrino field is charged in a noncommutative analogue of the adjoint representation with the matrix multiplication replaced by the $\star$-product. From a geometric point of
view, the interaction is seen as a modified photon-$\theta$ background
throughout which neutrinos tend to propagate.

All NC fields in the action are composite functions of commutative fields
(denoted as $A^\mu,\Psi$) expanded/expressed in powers of ordinary gauge
field via the $\theta$-exact SW maps
\begin{equation}
\begin{split}
\widehat A_\mu&=A_\mu-\frac{e\kappa}{2}\theta^{ij}A_i\star_2(\partial_jA_\mu+F_{j\mu})+\mathcal O(A^3)\,,
\\
\widehat\Psi&=\Psi-e\kappa\theta^{ij}A_i\star_2\partial_j\Psi+\mathcal
O(A^2)\Psi\,.
\end{split}
\label{exactPsi}
\end{equation}
This makes theory exact with respect to the noncomutative parameter $\theta$.  Here $\Psi$ means commutative $\Psi_{L \choose R}$, i.e.  left/right Dirac-type\footnote{Note that instead of SW map of Dirac neutrinos $\Psi$ one may consider a {\it chiral} SW map, which is compatible with grand unified models having chiral fermion multiplets \cite{Aschieri:2002mc}.} massive neutrino field,
and the $\star_2$-product is defined as follows:
\begin{equation}
f(x)\star_2 g(x)=f(x)\frac{\sin\frac{{\partial_x}\Lambda{\partial_y}}{2}}{\frac{{\partial_x}\Lambda{\partial_y}}{2}}g(y)\Bigg|_{y=x}.
\label{star2}
 \end{equation}
Expanding the action in terms of the commutative gauge fields, and isolating cubic terms up to the $A_\mu$ first order in Lagrangian, one obtains the relevant
$\theta$-exact Feynman rules:
\begin{equation}
\Gamma^{\mu}= ie\kappa \Big[(\slashed p - m) {(\theta q)}^{\mu}-(p\theta q)\gamma^{\mu} -{(\theta p)}^{\mu}{\slashed q}\Big]F(q,p),
\label{Feynrule}
\end{equation}
involving the following function $F(q,p)$,
\begin{equation}
F(q,p)=\frac{\sin \frac{q\theta p}{2}}{\frac{q\theta p}{2}},\;\;q\theta p \equiv q_i \theta^{ij} p_j\,.
\label{Fqk}
\end{equation}

As described in details in~\cite{Horvat:2011qn}, the coupling \eqref{Feynrule}
with one arbitrary $\kappa$ can be included into neutrino-mass extended noncommutative  standard model(s). It is also demonstrated in the section 2 of~\cite{Horvat:2011qn} that  different $\kappa$ values (or, more generally, different left/right charge combinations) can be assigned to different generations of matter fields which are minimally coupled to a U(1) gauge field via NC covariant derivative(s). On the other hand, following the analysis in the later sections of~\cite{Horvat:2011qn} one can easily notice that such generation dependence of $\kappa$ is, within the context of neutrino-mass extended noncommutative standard model(s), constrained by the gauge invariance of NC mass and/or Yukawa terms: Gauge invariance of a mass term $\bar N_1\star M_{12}\star N_2$ or a Yukawa term $\bar N_1\star H_{12}\star N_2$ requires the very left gauge transformation of $\bar N_1$ and the very right transformation of $N_2$ to cancel each other via the cyclicity of the Moyal star product~\cite{Horvat:2011qn}, which forces $N_1$ and $N_2$ to share the same $\kappa$-value for gauge invariance. From this viewpoint a universal $\kappa$-value across all flavor generations as in~\cite{Horvat:2011qn} actually allows most general neutrino mixing in the gauge invariant mass/Yukawa term constructions. Hence in the rest of the paper we will deal with universal but otherwise arbitrary $\kappa$ parameter.

Using standard techniques the plasmon decay rate into neutrinos (per generation) can be calculated to be \cite{Horvat:2011iv}
\begin{equation}
 {\Gamma_{\rm NC}(\gamma_{pl.} \to \bar\nu_{L \choose R}
\nu_{L \choose R})} = \kappa^2 \frac{\alpha}{2} \,\omega_{pl} \left(1-\frac{\sin
X}{X}\right)\,
\label{NCrate}
\end{equation}
with $\alpha$ being the fine structure constant. For the light-like noncommutativity preserving unitarity\footnote{The light-like case \cite{Aharony:2000gz} with notations $\theta^2 =(\theta^2)^{\mu}_{\mu} = \theta_{\mu\nu}\theta^{\nu\mu}
= 2(\vec{E}_{\theta}^2 - \vec{B}_{\theta}^2 )$ specified in \cite{Horvat:2012vn}, corresponds to
$|\vec{E_\theta}|=|\vec{B_\theta}|=1/(2\Lambda_{\rm{NC}}^2)$, and $\vec{E_\theta}\cdot\vec{B_\theta}=0$.} the full noncommutative effect will be still exhibited through $X={\omega_{pl}^2}/{(2\Lambda_{\rm NC}^2)}$.

It is important to note that the plasma frequency $\omega_{pl}$ is determined as the frequency of plasmons at $|\vec{q}| = 0$.  In the very high temperature regime,  where the mass of background electrons is irrelevant and can be put to zero, the dispersion relation for transverse and longitudinal waves can be calculated analytically, giving \cite{Grasso:1993bp}
\begin{equation}
\omega^2_{pl} = {\cal R}e \,\Pi_{(T/L)}(q_{0}, |\vec{q}| = 0) = \frac{e^2T^2}{9},
\label{plasdisprel}
\end{equation}
where ${\cal R}e \,\Pi_{(T/L)}$ is the transverse/longitudinal part of the one-loop contribution to the photon self-energy at finite temperature/density.

Now we continue with the investigation of the cosmic neutrino background in NC spacetimes.  The RH neutrino is commonly considered to decouple at the temperature $T_{dec}$ when the condition
\begin{equation}
{\Gamma(\gamma_{pl.} \to \bar\nu_{R} \nu_{R})}\simeq H(T_{dec}),
\label{decouple}
\end{equation}
is satisfied.  In this case the plasma frequency reads
\begin{equation} \omega_{pl}= \frac{eT_{dec}}{3}\,\sqrt{g^{\rm ch}_{*}},
\label{disprel}
\end{equation}
where $g^{\rm ch}_{*}$ counts all (effectively massless) charged-matter loops in $\Pi_{T/L}$. On the other hand, the Hubble parameter is given by
\begin{equation}
H(T_{dec}) =
\left(\frac{8 \pi^3}{90} g_{*}(T_{dec}) \right)^{1/2} \frac{T_{dec}^2}{M_{Pl}}\,,
\label{HT}
\end{equation}
and $g_*$ counts the total number of effectively massless degree of freedom. Further on, we stick with parameters $g_{*}$ and $g^{\rm ch}_*$ fixed at their SM values, $g_{*}\simeq g^{\rm ch}_*\simeq 100$.

Computing the decoupling temperature $T_{dec}$ based on the assumption that the decay rate (\ref{decouple}) is solely due to NC effects and comparing with lower bounds on $T_{dec}$ that can be inferred from observational data, we are now in position to determine lower bounds on the scale of noncommutativity $\Lambda_{\rm NC}$. Proceeding in this spirit, one finds that BBN provides the following relation between the decoupling temperature $T_{dec}$ and the NC scale
$\Lambda_{\rm NC}$:
\begin{equation}
\begin{split}
&T_{dec} \simeq \frac{\kappa^2}{2\pi}\sqrt{{5\alpha^3}\frac{g^{\rm ch}_*}{g_*}}M_{pl}\left(1-\frac{\sin X}{X}\right),
\\&
X=\frac{2\pi\alpha g^{ch}_* T_{dec}^2}{9\Lambda_{\rm NC}^2}\,.
\end{split}
\label{Tdecxi}
\end{equation}
Note that with fixed $g_{*}$ and $g^{\rm ch}_*$ one cannot simply dial down $\Delta N_{\nu}^{max}$ to arbitrary precision to accommodate $T_{dec}$ being proportional $M_{pl}$, as given by (\ref{Tdecxi}).  On the other hand, sensitivity to PTOLEMY requires small $T_{dec}$, which one can only achieve for $\left(1-\frac{\sin X}{X}\right)\ll 1$. This only occurs when $X\ll 1$, so in this limit we can use the leading order term in the expansion in $X$ to obtain:
\begin{equation}
\Lambda^4_{\rm
NC}\simeq\frac{\kappa^2\pi}{243}\sqrt{5\alpha^7(g^{\rm ch}_*)^5 g^{-1}_{*} }
\;M_{pl}\;T_{dec}^3\,.
\label{LaNC}
\end{equation}

Now setting $g_* =g^{\rm ch}_* = 100$, $M_{pl}=1.221\times 10^{19}$ GeV and
$T_{dec}\gtrsim200$ MeV (quark-hadron phase transition), a lower bound on
$\Lambda_{\rm NC}$ can be obtained as
\begin{equation}
\Lambda_{\rm
NC}\gtrsim 0.98\sqrt{\kappa}\; \rm TeV.
\label{8.35}
\end{equation}
For $T_{dec}\gtrsim200$
GeV ($EW$ phase transition), we have
\begin{equation}
 \Lambda_{\rm
NC}\gtrsim 175\sqrt{\kappa} \; \rm TeV.
 \label{10^3}
 \end{equation}
This bounds appear to be relatively mild in comparison with other similar bounds
~\cite{Horvat:2011iv,Horvat:2011qn,Horvat:2010sr,Horvat:2012vn}.  Also,
as shown below and in contrast with those lower bounds on $\Lambda_{\rm NC}$,  the  full numerical solution to \eqref{Tdecxi} will feature a maximal allowable
NC scale $\Lambda^{max}_{\rm NC}$, above which the RH neutrino can never stay in the thermal equilibrium via the NC coupling to photon and thus have no impact on PTOLEMY capture rate.

\begin{figure}[t]
\begin{center}
\includegraphics[width=8.7cm,angle=0]{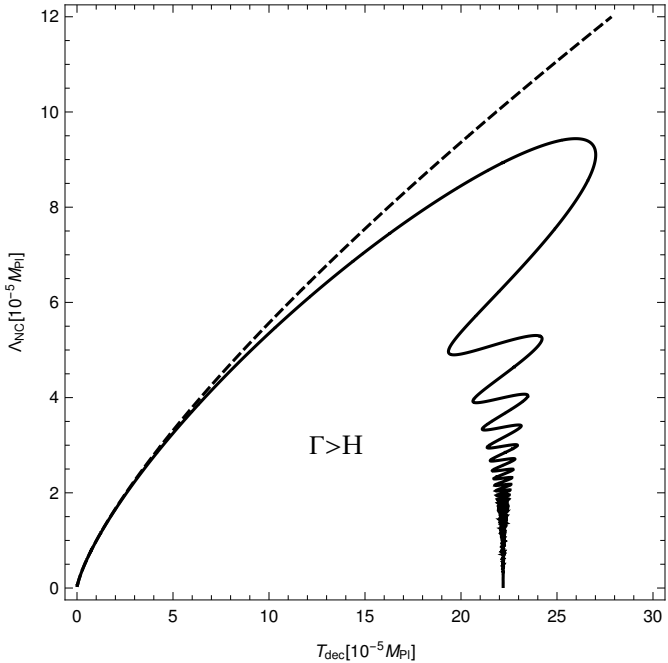}
\end{center}
\caption{Numerical plot of the scale of noncommutativity $\Lambda_{\rm NC}$ versus decoupling temperature $T_{dec}$ according to the eq.~\eqref{Tdecxi} (solid curve), and its $\theta$-first order approximation~\eqref{LaNC} (dashed curve).
In this plot we are using $\kappa=1$, and $g_* =g^{\rm ch}_* = 100$, respectively.}
\label{fig:Plasmon2}
\end{figure}

Since the equation~\eqref{Tdecxi} is exact with respect to the scale of
noncommutativity and decoupling temperature, it is interesting to extend our investigation to a temperature range well beyond the validity of the $\theta$-first order approximation~\eqref{LaNC}, which is done by numerical evaluation and shown in the Fig.~\ref{fig:Plasmon2}.  We find, surprisingly, that due to the
switch in the behavior of the plasmon decay rate from $~T^5$ at low
temperatures to $~T$ at very high temperatures the solution curve actually
drops down at a temperature range roughly independent from the NC scale and
singles out a closed region on the scale of noncommutativity $\Lambda_{\rm
NC}$ versus decoupling temperature $T_{dec}$.  Within this region surrounded
by the solid curve, the Hubble expansion rate~\eqref{HT} is always smaller than the
NC plasmon decay rate~\eqref{NCrate}. Therefore the higher temperature solution at each given noncommutative scale, sitting on the right-hand side of the solid curve, may be interpreted as the {\it coupling} temperature, i.e.  the temperature where the NC plasmon decay rate first time catches (or it may be the reheating temperature, whichever is lower) the Hubble rate during cooling of the universe after the Big Bang.

The appearance of a closed region where $\Gamma > H$ implies that the NC
scale can be bounded from above at $\Lambda_{\rm NC}^{max} \simeq 0.95 \times 10^{-4} M_{Pl}$. For NC scales $\Lambda_{\rm NC}>\Lambda_{\rm NC}^{max}$ RH neutrinos stay out of thermal equilibrium at any temperature. For each NC scale $\Lambda_{\rm NC}<\Lambda_{\rm NC}^{max}$, there exists two temperature scales, namely a, lower, decoupling temperature $T_{dec}$ and a higher, coupling
temperature $T_{couple}$. As a consequence, RH neutrinos can only stay in effective thermal contact with the rest of the universe for the temperature range
$T_{dec}\le T\ge T_{couple}$.\footnote{Actually neutrinos could enter and exit thermal equilibrium a few times beneath the high temperature boundary $T_{couple}$ due to the oscillatory nature of the NC production rate.} In other words, during cooling of the universe, RH neutrinos first time enter thermal equilibrium when temperature reaches $T_{couple}$. As the temperature decreases further, the decay rate, starting at $T_{dec}$, drops once again below the Hubble rate and sterile neutrinos finally decouple.

\begin{figure}[t]
\begin{center}
\includegraphics[width=8.7cm,angle=0]{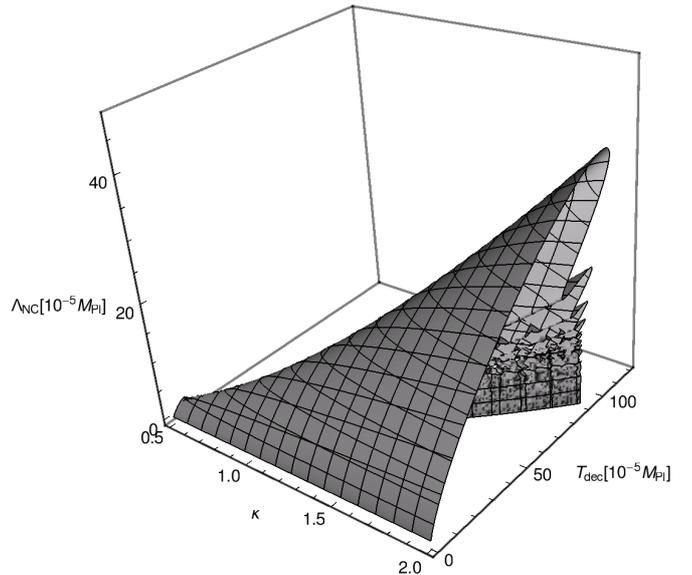}
\end{center}
\caption{3D plot of the decoupling relation~\eqref{Tdecxi} with respect to temperature, NC scale $\Lambda_{\rm NC}$, and coupling ratio $\kappa$ range $[0.5,2]$ for degrees of freedoms $g_*=g^{\rm ch}_*=100$. The high temperature boundary of the plot decays very quickly due to the $\kappa^2$ dependence in $T_0$~\eqref{T_0}.}
\label{fig:Plasmon3}
\end{figure}

The equation~\eqref{Tdecxi} allows us to estimate the bound on $T_{couple}$ analytically: In any case it has to be smaller than a fixed temperature scale
\begin{equation}
T_0=\frac{\kappa^2} {2\pi}\sqrt{5\alpha^3\frac{g^{\rm ch}_*}{g_*}}M_{pl}\Bigg|_{\kappa=1} \simeq
2.22\times 10^{-4}\kappa^2  M_{pl},
\label{T_0}
\end{equation}
multiplying the maximum value ($\simeq 1.217$) of the $(1-\sin X/X)$ term sitting in the parenthesis, while for sufficiently small NC scales $T_{couple}$ converges to $T_0$. These facts provide an estimation for $T_{couple}$'s maximal value $T^{max} \simeq 1.22 T_0\simeq 2.7 \times 10^{-4}\kappa^2 M_{Pl}$.  Via $T_0$, $T^{max}$ depends on the quadratic power of the parameter $\kappa$ and gets suppressed rather quickly when $\kappa$ decreases, as illustrated in 3D Fig.~\ref{fig:Plasmon3}.

The existence of $T_{couple}$, bounded from above by $T^{max}\simeq 2.7 \times 10^{-4} M_{Pl}$ and an upper bound on the scale of noncommutativity $\Lambda_{\rm NC}^{max} \simeq 0.95 \times 10^{-4} M_{Pl}$ for RH neutrino to reach thermal equilibrium via NC coupling to photon from Fig.~\ref{fig:Plasmon2} represent additional results of our work. 
Now we note that decoupling of the production rate for two branches of  $T_{dec}$ ($T_{dec}$ and $T_{couple}$) exhibits certain similarity to the UV/IR mixing in the radiative corrections of the NC theories. Both phenomena share the same origin from the exact/nonperturbative treatment of the NC parameter $\theta$ in the quantum theory as well.

In total we have shown that the PTOLEMY total capture rate in the Dirac
neutrino case may be enhanced in the present scenario up to 20$\%$ (10$\%$)
if the scale of noncommutativity $\Lambda_{\rm NC} \gtrsim {\cal O}(1)$
TeV $(\gtrsim {\cal O}(100))$ TeV. This is still consistent with the bunch of constraints on the scale of noncommutativity obtained from particle physics phenomenology \cite{Chaichian:2001py,Hinchliffe:2002km,Ohl:2004tn,Alboteanu:2006hh,Ettefaghi:2007zz,Alboteanu:2007bp,Alboteanu:2007by}. If, however, one adopts a more ``natural'' value for $\Lambda_{\rm NC}$, which is closer to the string (or even Planck \cite{Abel:2006wj}) scale, then the total capture stays as predicted by the standard theory.  Hence, the results of the PTOLEMY experiment could not only be used as a test of noncommutative gauge field theories, but also could provide an independent constraint on the scale of noncommutative deformation of spacetime as well.

\section*{Acknowledgment} We would like to thank to G.Y.  Huang for drawing
our attention on refs \cite{Zhang:2015wua,Huang:2016qmh} considering the
PTOLEMY experiment, and for encouraging comments regarding our work.  This
work is supported by the Croatian Science Foundation (HRZZ) under Contract
No.  IP-2014-09-9582, and by the COST Action MP1405 (QSPACE).  J.Y.
acknowledge support by the H2020 Twining project No.  692194,
"RBI-T-WINNING".  J.T.  and J.Y.  would like to acknowledge support of W.
Hollik, and MPI Munich, for hospitality.

\end{document}